\documentclass[a4paper,11pt]{article}
\usepackage{pos}

\newcommand{\figsize}{0.6}
\newcommand{\figsizetwo}{0.49}
\newcommand{\msub}[1]{\ensuremath _{\mbox{\scriptsize #1}}} 
\newcommand{\nii}{n\msub{i}}
\newcommand{\naa}{n\msub{a}}

\title{High temperature $U(1)_A$ breaking in the chiral limit}

\author*[]{Tam\'as G.\ Kov\'acs}

\affiliation[]{Department of Physics and Astronomy, ELTE E\"otv\"os Lor\'and
  University \\ 
  P\'azm\'any P\'eter s\'et\'any 1/a, H-1117 Budapest, Hungary}

\emailAdd{tamas.gyorgy.kovacs@ttk.elte.hu}

\abstract{We solve the long-standing problem concerning the fate of the chiral
  $U(1)_A$ symmetry in QCD-like theories at high temperature in the chiral
  limit. We introduce a simple instanton based random matrix model that
  precisely reproduces the properties of the lowest part of the lattice
  overlap Dirac spectrum. We show that in the chiral limit the instanton gas
  splits into a free gas component with a density proportional to $m^{N_f}$
  and a gas of instanton-antiinstanton molecules. While the latter do not
  influence the chiral properties, for any nonzero quark mass the free gas
  component produces a singular spectral peak at zero that dominates
  Banks-Casher type spectral sums. By calculating these we show that the
  difference of the pion and delta susceptibility vanishes only for three or
  more massless flavors, however, the chiral condensate is zero already for two
  massless flavors.  }

\FullConference{The 40th International Symposium on Lattice Field Theory
  (Lattice 2023)\\ 
July 31st - August 4th, 2023\\
Fermi National Accelerator Laboratory\\}

\begin{document}
\maketitle

\section{Introduction}

The approximate $SU(2)_A \times U(1)_A$ chiral symmetry of light quarks is an
essential feature of quantum-chromodynamics (QCD). While the $SU(2)_A$ part,
spontaneously broken at low temperatures, is expected to be restored above the
crossover temperature $T_c$, the flavor singlet $U(1)_A$ part is anomalous due
to the presence of instantons. The fate of the flavor singlet part is the
subject of a long debate that was started by the seminal paper of Pisarski and
Wilczek, who discussed possible scenarios for the finite temperature chiral
transition in the chiral limit \cite{Pisarski:1983ms}. Their results were
based on the $\epsilon$-expansion, and there has been an ongoing effort to
check it using lattice QCD.

\begin{figure}[t]
  \centering
  \includegraphics[width=\figsize\textwidth]{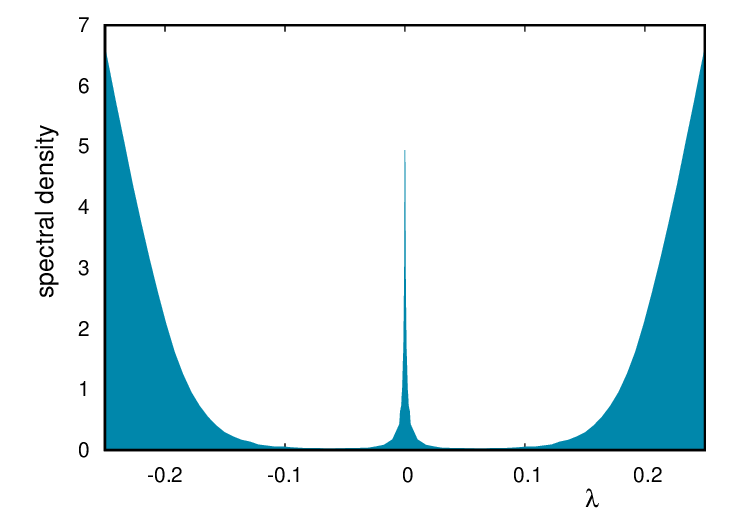}
  \caption{The distribution of the lowest Dirac eigenvalue on configurations
    containing exactly one instanton-antiinstanton pair on the lattice and in
    the random matrix model in a spatial volume of $32^3$ in lattice units
    (left panel). The same comparison for the second lowest eigenvalue on a
    larger volume of $40^3$ (right panel). For comparison we also show the
    distribution used for the fit, the one in the previous figure.}
  \label{fig:spd_b6.13}
\end{figure}

In the chiral limit the order parameter of the spontaneously broken chiral
symmetry, $\langle \bar{\psi} \psi \rangle$ can be written in terms of the
spectral density $\rho(\lambda)$ of the quark Dirac operator as
$$
  \lim_{m\rightarrow 0} \langle \bar{\psi} \psi \rangle \propto \rho(0).
$$
This is the Banks-Casher relation \cite{Banks:1979yr}, showing that the low
lying spectrum of the Dirac operator is intimately related to how chiral
symmetry is realized. Indeed, for some time the standard lore was that at the
transition temperature to the quark-gluon plasma, the spontaneously broken
chiral symmetry is restored, and $\rho(0)$ vanishes, up to small explicit
breaking effects due to the finite, but small light quark masses.

However, this view was challenged when instead of a small $\rho(0)$, a sharp
rise was found in the spectral density of the overlap Dirac operator at zero
\cite{Edwards:1999zm}. The reason this spike in the spectral density went
unnoticed before, was that unlike the overlap, the staggered and Wilson lattice
Dirac operators, used exclusively previously, did not respect chiral symmetry,
and could not properly resolve the smallest Dirac eigenvalues, the ones that
make up the spike of the density. For some time this finding remained largely
ignored, mostly because it was considered to be a quenched artifact, the
result of ignoring the quark determinant in the path integral. Indeed, the
determinant disfavors eigenvalues of small magnitude, and is expected to
suppress the spike in the spectral density at zero. Later on, the spike was
found to persist even in the presence of light dynamical quarks
\cite{Alexandru:2015fxa,Kaczmarek:2021ser,Ding:2020xlj,Meng:2023nxf}.
However, some doubts could still remain, as the lattice fermions used in these
works were not chirally symmetric, and their poor resolution of the small
Dirac eigenvalues might have lead to an improper suppression of the
spike. Indeed, results by the JLQCD collaboration using chiral lattice
fermions suggest that toward the chiral limit the spectral spike completely
disappears already at a nonzero quark mass
\cite{Aoki:2020noz,Aoki:2021qws}. Another possibility, the one we advocate in
the present paper, is that indeed, chiral fermions are needed for a proper
suppression of the spike, but the spike is still present at any finite quark
mass, however, it becomes undetectably small with the currently used lattice
volumes and statistics.

It seems to us that using the currently available lattice technology it is not
possible to explore the fate of the spike in the spectral density and also the
fate of the $U(1)_A$ symmetry for light dynamical quarks. In the present work
we suggest a different approach, based on the finding that in the quenched
theory the statistics of the eigenvalues in the spectral spike are consistent
with mixing zero modes of a free instanton gas, a proposal already put forward
in Ref.~\cite{Edwards:1999zm} and recently confirmed in more detail
\cite{Vig:2021oyt}.

\section{Random matrix model}

We propose a random matrix model for the description of the mixing zero
modes, the zero mode zone (ZMZ) of the free instanton gas in quenched QCD. In
a free instanton gas the number distribution of instantons and antiinstantons
are independent and identical Poisson distributions with mean $\chi_0 V/2$,
where $\chi_0$ is the topological susceptibility. A random matrix is
constructed by first drawing the number of instantons $\nii$ and
antiinstantons $\naa$ from the Poisson distributions. The size of the matrix
is $(\nii+\naa) \times (\nii+\naa)$. Since zero modes of like charge objects
are protected from mixing by the index theorem, the matrix has two diagonal
blocks of zeros of size $\nii \times \nii$ and $\naa \times \naa$. At high
temperatures the instanton zero mode wave functions decay exponentially with
exponent $\pi T$ \cite{Gross:1980br}, and we assume that the mixing matrix
elements of instanton and antiinstanton zero modes also follow this
exponential decay with the distance of the given instanton and
antiinstanton. In a noninteracting gas, the location of the topological
objects are chosen randomly within a three-dimensional\footnote{At high
  temperature the instantons typically occupy the whole available space in the
  temporal direction, so we ignore the temporal location of the instantons.}
box of size $L$. In this way the remaining off-diagonal blocks of the matrix
are filled with elements of the form $w_{kl} = A\cdot \exp(-\pi T r_{kl})$,
where $r_{kl}$ is the distance of the randomly placed instanton $k$ and
antiinstanton $l$. This completes the construction of one random matrix, and
we can easily generate ensembles of such random matrices.

\begin{figure}[t]
  \centering
  \includegraphics[width=\figsizetwo\textwidth]{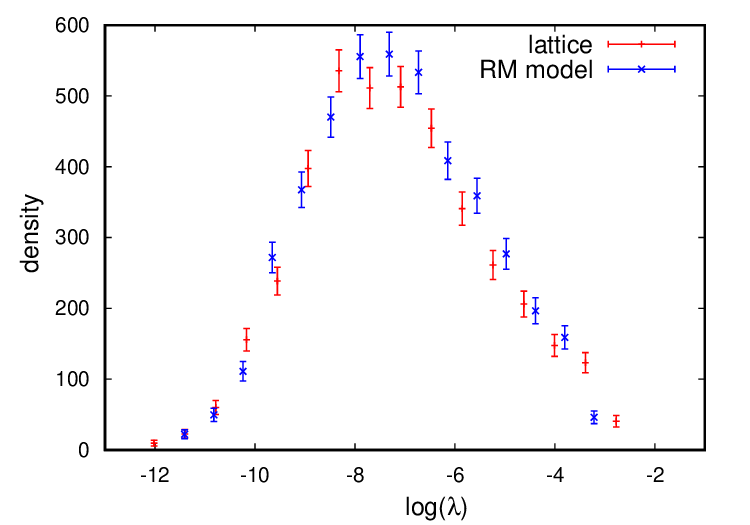} \hfill
  \includegraphics[width=\figsizetwo\textwidth]{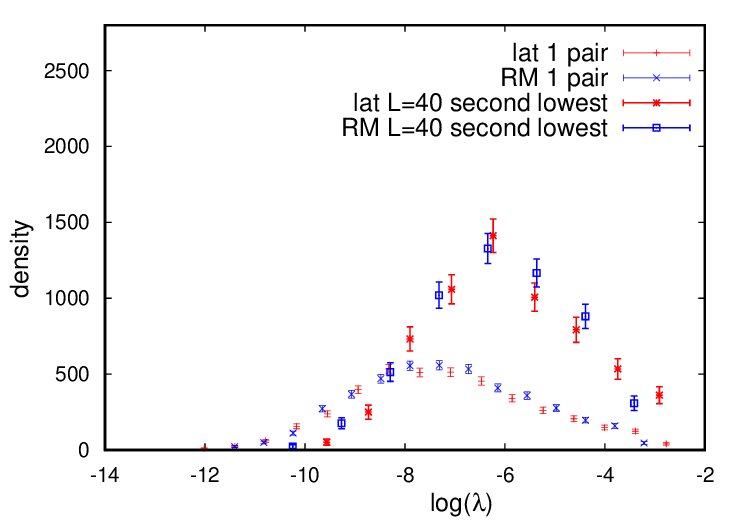}
  \caption{The distribution of the lowest Dirac eigenvalue on the lattice
    ensemble described in the text, compared to that in the matrix model with
    the best fit parameter $A$ (left panel). In both cases we used only the
    configurations with exactly one instanton-antiinstanton pair , and to
    better resolve the spike in the spectral density, we plotted the
    distribution of the natural log of the eigenvalues. The right panel shows a
  similar comparison, but with the second lowest eigenvalue, and on a larger
  lattice with $L=40$.}
  \label{fig:1pair}
\end{figure}

The model has two parameters, the topological susceptibility $\chi_0$ and the
prefactor $A$. To determine these parameters we consider an ensemble of 20k
quenched lattice configurations generated with the Wilson action at
$T=1.11T_c$ on lattices of size $32^3\times 8$. Computing the lowest
eigenvalues of the overlap Dirac operator on this ensemble and counting the
exact zero eigenvalues yields the topological susceptibility. For fitting the
single remaining parameter $A$, we consider the distribution of the lowest
nonzero Dirac eigenvalue on those configurations that have exactly two complex
conjugate eigenvalues in the ZMZ, corresponding to an instanton-antiinstanton
pair. In Fig.~\ref{fig:1pair} (left panel) we compare the distribution of the
lattice eigenvalues with that of the random matrix model with the best fit
parameter $A$. It is already remarkable that the whole distribution can be
fitted with just this one parameter, but now that the model is fixed we can
compare different properties of the Dirac spectrum on the lattice and in the
matrix model. The comparison can be made also for the distribution of the
second lowest eigenvalue or the full spectral density, and also on different
volumes. As an illustration, in Fig.~\ref{fig:1pair} (right panel) we show the
comparison for the second lowest eigenvalue on a larger volume. These tests
demonstrate that the random matrix model properly describes the details of the
lattice overlap spectrum. Simulations of the random matrix model on larger
volumes, not accessible to direct lattice simulations, indicate the in the
thermodynamic limit the spectral density is singular at zero
\cite{Kovacs:2023vzi}.

\section{Including dynamical quarks}

On the lattice, including dynamical quarks means that in addition to the
quenched Boltzmann weights, each configuration gets another weight factor
proportional to the quark Dirac determinant $\det(D+m)^{N_f}$. For simplicity
we assume $N_f$ degenerate quark flavors. The determinant of the lattice Dirac
operator can be split into the contribution of the zero mode zone and that of
the bulk as
\begin{equation}
  \det(D+m) = \prod_{i \in ZMZ} (\lambda_i +m) \cdot
  \prod_{i \in \mbox{\tiny bulk}} (\lambda_i +m).
\end{equation}
As can be seen in Fig.~\ref{fig:spd_b6.13}, at high temperatures the ZMZ and
the bulk are well separated, therefore eigenvalues in the bulk are not
expected to be correlated with the ones in the ZMZ. Our main concern here is
to study how the determinant suppresses the eigenvalues in the ZMZ. It is thus
a good approximation to ignore the bulk contribution to the determinant, as
that will only give a trivial factor in the path integral for quantities
depending on the ZMZ. Especially for small quark mass, the suppression of the
eigenvalues in the ZMZ will be driven by the contribution of those small
eigenvalues to the determinant. This is exactly the part of the Dirac spectrum
that is faithfully represented by the random matrix model, so including the
determinant of the random matrices in the statistical weight will properly
describe the suppression of the spectral spike by dynamical quarks. Thus the
random matrix model we propose for the ZMZ of QCD with dynamical quarks is the
one described in the previous section, with the additional weight
$\det(D+m)^{N_f}$ for each instanton configuration, where $D$ is the random
matrix corresponding to the given instanton configuration.

The numerical simulation of this model reveals that the topological
susceptibility is suppressed by light dynamical quarks as
\begin{equation}
  \chi(m) = \chi_0 m^{N_f},
  \label{eq:chi}
\end{equation}
where $\chi_0$ is the quenched susceptibility, the one obtained without
including the quark determinant. To understand this behavior we note that
throughout the simulations we found that the eigenvalues in the spectral spike
always satisfied $|\lambda| \ll m$, even for the smallest quark mass of
$m=0.01$ (in lattice units). The reason for this is that as the quarks become
lighter, the determinant suppresses the number of instantons, the typical
instanton-antiinstanton distance grows, and the matrix elements become
exponentially small, resulting in ever smaller eigenvalues. As a result, in
the chiral limit the magnitude of the eigenvalues in the ZMZ decreases much
faster than the quark mass, and they will always remain smaller than the quark
mass. If the eigenvalues are much smaller than the quark mass then to a very
good approximation the determinant of a matrix with $\nii$ instantons and
$\naa$ antiinstantons can be written as
\begin{equation}
   \det(D+m) = \prod_i (\lambda_i+m) \approx m^{\nii+\naa}.
\end{equation}

\begin{figure}[t]
  \centering
  \includegraphics[width=\figsizetwo\textwidth]{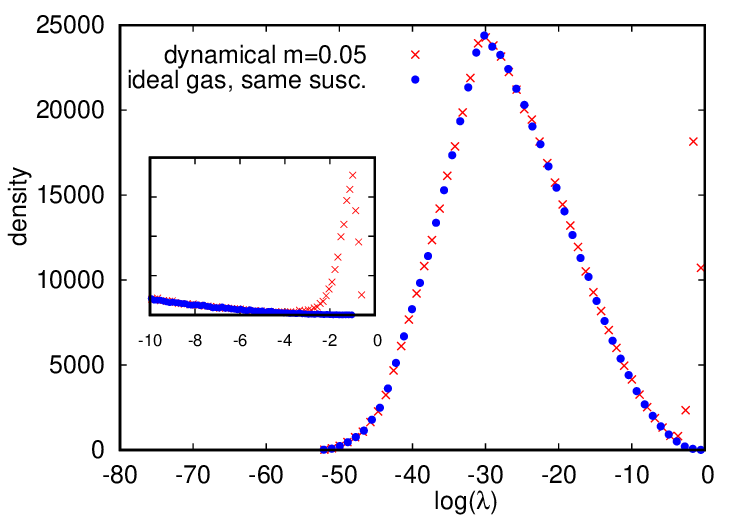} \hfill
  \includegraphics[width=\figsizetwo\textwidth]{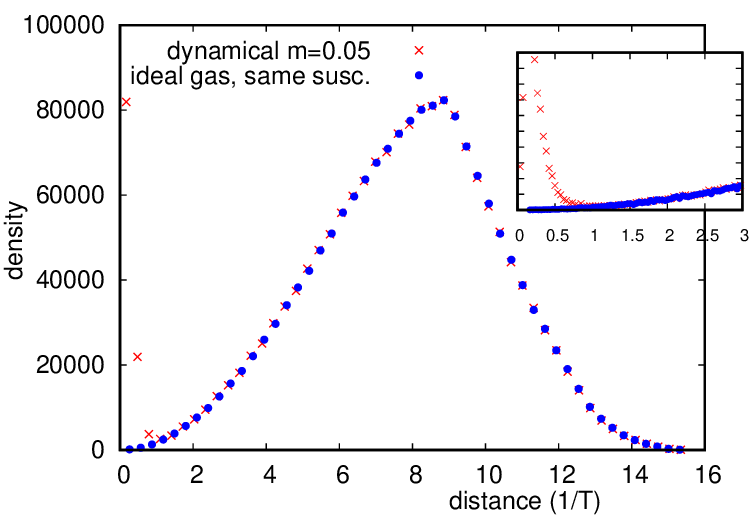}
  \caption{The spectral density of the random matrix model with two degenerate
    dynamical quark flavors of mass $m=0.05$ (in lattice units) compared to
    the spectral density of the matrix model of the free instanton gas with
    the same topological susceptibility (left panel). A comparison of the
    density of nearest opposite charged topological objects between the
    quenched and the dynamical random matrix ensemble (right panel).}
  \label{fig:spd_m0.05}
\end{figure}

With $N_f$ quark flavors, each (anti)instanton contributes a suppression
factor $m^{N_f}$ to the determinant, and the distribution of (anti)instanton
numbers is still Poissonian, but with a susceptibility suppressed as
$\chi_0 \rightarrow \chi_0 m^{N_f}$. This explains the quark mass dependence
of the susceptibility in Eq.~(\ref{eq:chi}). The fact that the Poisson
distribution of the number of topological objects is preserved also implies
that even in the presence of light dynamical quarks, the lowest part of the
Dirac spectrum can still be understood as the zero mode zone of a free
instanton gas. To demonstrate this, in the left panel of
Fig.~\ref{fig:spd_m0.05} we compare the spectral density of the random matrix
model with two degenerate dynamical quark flavors of mass $m=0.05$ (in lattice
units) with that of the matrix model of the free instanton gas (without the
determinant) with the same topological susceptibility. The two curves exactly
agree, except for the largest eigenvalues, where the model with dynamical
quarks shows an excess of eigenvalues. Large eigenvalues in the matrix model
indicate that there might be large matrix elements which in turn would imply
that in the dynamical case there are nearby instanton-antiinstanton pairs. To
check that, in the right panel of Fig.~\ref{fig:spd_m0.05} we compare the
density of the nearest opposite charged objects as a function of their
distance for the dynamical and quenched matrix ensembles of the left panel of
the same figure. Indeed, in the dynamical case we find an excess of tightly
bound instanton-antiinstanton pairs at a distance smaller than the instanton
size $1/T$. These instanton-antiinstanton ``molecules'' are held together by
the attractive force due to light dynamical quarks.

\section{Chiral condensate and $U(1)_A$ breaking susceptibility}

We have seen that with light dynamical quarks the instanton gas has two
components. There is a dilute gas of free instantons, responsible for the
small Dirac eigenvalues, i.e.\ the spectral spike. Besides that, there is a
component of tightly bound instanton-antiinstanton molecules. Our simulations
reveal that in the chiral limit the eigenvalues corresponding to these two
components behave differently. While the free instanton eigenvalues in the
spike become smaller as the gas becomes more dilute, the eigenvalues
corresponding to the molecules remain at a fixed scale in the spectrum. This
is because the size of the molecules does not change with the quark mass. An
important consequence of this is that in the chiral limit any nonvanishing
contribution to Banks-Casher type sums can only come from the free instanton
generated part of the spectrum.

The simplest of these sums is the one providing the chiral condensate. In the
chiral limit the chiral condensate can be written in terms of the Dirac
spectrum as
\begin{equation}
   \langle \bar{\psi} \psi \rangle \approx
     \langle \sum_i \frac{m}{m^2 + \lambda_i^2} \rangle
      \approx 
       \underbrace{\left( \mbox{\parbox{10ex}{\tiny avg.\ number
             of instantons in free gas}}\right) }_{\chi_0 m^{N\msub{f}} V}
     \cdot \frac{1}{m} \; = \; m^{N\msub{f}-1} \chi\msub{0} V.
  \label{eq:pbp}
\end{equation}
Here we used the fact that the eigenvalues corresponding to the molecular
component of the instanton gas remain at a fixed scale, and in the chiral
limit they do not contribute to the sum. In contrast, the magnitude of
eigenvalues generated by the free gas component becomes smaller in the chiral
limit so rapidly that $\lim_{m \rightarrow 0} \frac{\lambda}{m} =0$, and all
these eigenvalues will contribute a term $1/m$ to the sum. So the chiral
limit of the condensate is given by the contribution of the free instanton
gas. This result shows that for two and more flavors of vanishing mass, the
chiral condensate goes to zero, and the spontaneously broken chiral symmetry
is restored, as expected. We also note that Eq.~(\ref{eq:pbp}) is consistent
with the expansion of the QCD free energy density in terms of the quark mass
around the chiral point \cite{Kanazawa:2015xna}, and also consistent with the
resulting quasi-instanton picture of \cite{Kanazawa:2014cua}. In fact, our
instanton-based matrix model provides a physical explanation for the
quasi-instanton picture, and shows that it is valid not only in the small
quark mass limit.

Another quantity of interest in the chiral limit is the susceptibility
$\chi_\pi - \chi_\delta$, a nonzero value of which signals the breaking of the
$U(1)_A$ symmetry
\cite{Kaczmarek:2021ser,Ding:2020xlj,Aoki:2020noz}. Similarly to the chiral
condensate, in the chiral limit this quantity can also be written as a sum
over the Dirac spectrum as
\begin{equation}
\chi_\pi - \chi_\delta \approx
     \langle \sum_i \frac{m^2}{(m^2 + \lambda_i^2)^2} \rangle
     \approx 
      m^{N\msub{f}} \chi\msub{0} V
     \cdot \frac{1}{m^2} \; = \; m^{N\msub{f}-2} \chi\msub{0} V,
\end{equation}
showing that even for two flavors of vanishing mass, the symmetry remains
effectively broken. 

\acknowledgments

This work was supported by Hungarian National Research, Development and
Innovation Office NKFIH Grant No. KKP126769 and NKFIH excellence grant
TKP2021-NKTA-64.

\end{document}